\documentclass[amsmath,amssymb,aps,prb,citeautoscript,
superscriptaddress,twocolumn,showpacs,10pt]{revtex4-1}

\usepackage{graphicx}
\usepackage{dcolumn}
\usepackage{bm}

\usepackage{mathrsfs}
\usepackage[colorlinks,linkcolor=red,
anchorcolor=green,
citecolor=blue,
breaklinks]{hyperref}

\graphicspath{{figures/}}
\begin{document}

\title{Preferred States of Open Electronic Systems}

\author{HaoXiang Jiang}
\affiliation{Guangdong Provincial Key Laboratory of Quantum Engineering and Quantum Materials, School of Physics and Telecommunication Engineering, South China Normal University, Guangzhou, 510006, China}

\author{Yu Zhang}
\email{zhy@lanl.gov}
\affiliation{Physics and Chemistry of Materials, Theoretical Division, Los Alamos National Laboratory, Los Alamos, NM, 87545, USA}

\date{\today}

\begin{abstract}
System-environment interaction may introduce dynamic destruction of quantum coherence, resulting in the transition from superposed states to incoherent mixtures which have special representation named as pointer states. Here, pointer states of an open electronic system which is driven out of equilibrium are studied. The decoherence effect is taken into account through two different models which are B\"{u}ttiker's virtual probe model and electron-phonon interaction in the polaron picture. The pointer states of the system with different coupling strength are investigated. The pointer states are identified by tracking the eigenstates of the density matrix in real-time evolution. It is found that the pointer states can emerge for arbitrary coupling strength. And the pointer states deform to the eigenstates of the system in the strong coupling limit, which indicates the vanish of quantumness in the strong coupling limit. 
\end{abstract}

\pacs{03.65.Yz, 05.30.-d, 05.60.Gg}

\maketitle

\section{Introduction}
Bohr's explanation of quantum mechanics puts an unnatural boundary between the quantum and classical worlds. The emergence of classicality from the quantum may be explained by decoherence~\cite{PhysRevD.24.1516,PhysRevD.26.1862,zurek2009, RevModPhys.75.715, zurek2003arxiv,zurek1991}. The loss of quantum coherence of the system is induced by its interaction with the environment, which is universal because almost all the realistic physical systems can not be isolated from the surrounding environment. During the decoherence process, special states are selected, which are called ``preferred states" or ``pointer states"~\cite{PhysRevD.24.1516,PhysRevD.26.1862,zurek2009, RevModPhys.75.715}. In quantum transport, decoherence is crucial and responsible for the transition from ballistic transport to Ohm's law in nanoscale.

Pointer states have been studied in different models~\cite{PhysRevLett.108.070403,PhysRevA.89.014104,PhysRevE.86.021109, PhysRevLett.85.3552,PhysRevLett.86.2913,PhysRevLett.70.1187,PhysRevLett.82.5181,PhysRevA.77.012108,PhysRevLett.98.130401,PhysRevE.81.051127} by either analytical or numerical methods. The emergence of pointer states explains the transition from the quantum world to classical one due to a process of natural selection induced by the environment which is analogous to the Darwinian selection rule. This selection process is termed as quantum Darwinism~\cite{zurek2009}. The selection process, called einselection, destroys the superposition of pointer states through the continuous system-environment interaction. Thus, the vast potentiality of superposed states is suppressed, resulting in a reduced set of pointer states. All the system-environment interaction leads to decoherence in a particular basis determined by the physics of the interaction. Zurek and his collaborators have shown that a preferred basis which a quantum system will evolute into is the pointer basis underlying predictable classical states~\cite{zurek2009}. It is in this sense that the pointer states of classical reality are selected from the quantum counterpart and exist in the macroscopic realm in a state that can undergo further evolution.

Quantum transport through a certain system is fundamentally an open electronic system. The interplay of quantum interference and decoherence can significantly affect the performance of electronic devices~\cite{jz5007143}. The selection of pointer states is accompanied by the destruction of quantum interference. But little attention has been paid to study the pointer states in quantum transport. In this work, the pointer states of a molecular device are studied. According to the definition of pointer states, pointer states are the most robust to the system-environment interaction, while the superposition of these pointer states are quickly destroyed during the decoherence process. This is also regarded as the stability criterion for the selection of pointer states~\cite{PhysRevD.24.1516,PhysRevD.26.1862}. Detailed knowledge of pointer states of an open electronic system is also important for practical implementations. Because pointer states are robust to the environment, the system may be protected from the decoherence by designing the environment.

In this paper, pointer states of an open electronic system are investigated. Different from the previous models where only a single environment is involved, the system under study is connected to two fermionic baths which act as source and drain driving the system out of equilibrium. The pointer states of the system with Buttiker's virtual probe~\cite{PhysRevB.33.3020} or electron-phonon interaction are numerically examined. The Buttiker's virtual probe mode or electron-phonon interaction introduces the source of decoherence. It is found that the pointer states of the system deform to the eigenstates of the system in the strong coupling limit. The decoherence rate with different coupling strength is also studied and is found to decrease with increasing coupling strength. 

The paper is organized as follows: In Sec.~\ref{sec:method}, the methodology of treating time-dependent quantum transport is briefly introduced within the none-equilibrium Green's function formalism. The decoherence effect is taken into account in two different models. Sec.~\ref{sec:result} examines the pointer states of a benzene molecule sandwiched by two leads. The decoherence dynamics and evolution of pointer states in different coupling regime is illustrated with corresponding discussions. Finally, a summary is given in Sec.~\ref{sec:summary}.

\section{Model and methodology}\label{sec:method}
\subsection{Time-dependent quantum transport theory for non-interacting system}
The system considered here is a benzene molecule coupled to  two leads as illustrated by Fig.~\ref{figbenzene}. The two leads act as source and drain respectively. The total Hamiltonian is consisted of the molecule ($H_M$), leads ($H_\alpha, \alpha=L,R$) and the coupling between them $H_{M\alpha}$,
\begin{equation} 
H_T=H_M+\sum_\alpha(H_{M\alpha}+H_\alpha).
\end{equation}
The Hamiltonians of the molecule, leads and the coupling between them are expressed as
\begin{eqnarray}
H_M=&&\sum_\mu \epsilon_\mu d^\dag_\mu d_\mu+\sum_{\mu\neq\nu}
t_{\mu\nu}(d^\dag_\mu d_\nu+\text{H.c.}),
\nonumber\\
H_\alpha=&&\sum_{\alpha k_\alpha}\epsilon_{k_\alpha}
c^\dag_{k_\alpha}c_{k_\alpha},
\nonumber\\
H_{M\alpha}=&&
\sum_{\alpha k_\alpha \mu}
(V_{k_\alpha \mu}c^\dag_{k_\alpha} d_\mu+\text{H.c.}).
\end{eqnarray}
Where $d_\mu$ ($d^\dag_\mu$) and $c_{k_\alpha}$ ($c^\dag_{k_\alpha}$) are annihilation (creation) operators of electrons in molecule and lead $\alpha$, respectively. This system has been studied extensively in molecular electronics~\cite{PhysRevB.88.054301,ratner2013, aradhya2013,guedon2012, solomon2010,jz5007143}.  $\epsilon_\mu$ denotes the onsite energy of site $\mu$. $t_{\mu\nu}$ is the hopping parameter between site $\mu$ and $\nu$. Here, only the nearest site hopping is considered. $\epsilon_{k_\alpha}$ is the energy of $k_\alpha$-th electronic state in lead $\alpha$. The interaction between the molecule and lead $\alpha$ is characterized by the coupling strength $V_{k_\alpha\mu}$. Due to the coupling to leads, electronic states of the molecule are renormalized, which is represented by the self-energy or line-width function. The line-width function of lead $\alpha$ is given by 
\begin{equation}
\Gamma_{\alpha,\mu\nu}(\epsilon)=2\pi \sum_{k_\alpha}
V^*_{k_\alpha \mu}V_{k_\alpha \nu}
\delta(\epsilon-\epsilon_{k_\alpha}).
\end{equation}
where $\delta(\epsilon-\epsilon_\alpha)$ is the density of states of lead $\alpha$. In this study, the lead is modeled by a tight-binding chain with internal hopping parameter $t_\alpha$, the the line-width function is obtained as
\begin{equation}
\Gamma_{\alpha,\mu\nu}(\epsilon)=V_{\alpha,\mu}V_{\alpha,\nu}
\frac{\sqrt{4t^2_\alpha-(\epsilon-\mu_\alpha)^2}}{t^2_\alpha}.
\end{equation}
where $\mu_\alpha$ is the chemical potential of lead $\alpha$ and $V_{\alpha,\mu}$ is the coupling strength between site $\mu$ and lead $\alpha$. If wide-band limit (WBL) approximation~\cite{PhysRevB.50.5528} is adopted, the line-width function becomes $\Gamma_{\alpha,\mu\nu}=V_{\alpha,\mu}V_{\alpha,\nu} \frac{2}{t_\alpha}$. Since this work only considers the nearest site coupling, only one site is coupled to each lead. Consequently, $\Gamma_\alpha$ contains only one nonzero value, which is represented as $\Gamma$.

The evolution of reduced single-particle density matrix (RSDM) is solved through equation of motion (EOM) methods,
\begin{equation}
i\dot{\sigma}(t)=[h(t),\sigma(t)]-
\sum_\alpha [\varphi_\alpha(t)-\varphi^\dag_\alpha(t)],
\end{equation}
where $\sigma(t)$ is the RSDM and $\varphi_\alpha(t)$ is auxiliary density matrix, which accounts for the interaction between the molecule and leads.  Within WBL approximation and Pad\'{e} expansion of Fermi function, $\varphi_\alpha(t)$ reads~\cite{PhysRevB.87.085110},
\begin{eqnarray}
\varphi_\alpha(t)=&& i[\sigma(t)-1/2]\Lambda_\alpha
+\sum^N_k \varphi_{\alpha k}(t),\nonumber\\
i\dot{\varphi}_{\alpha k}(t)=&&
-\frac{2i\eta_k}{\beta}\Lambda_\alpha -
[\epsilon_{\alpha k}(t)-h(t)+i\Lambda]\varphi_{\alpha k}(t),
\end{eqnarray}
where $\Lambda=\sum_\alpha \Lambda_\alpha$ is the total line-width function, and 
\[
\epsilon_{\alpha k}(t)=p_k+\Delta_\alpha(t)
\]
where $\Delta_\alpha(t)$ is the time-dependent voltage applied on lead $\alpha$. $p_k$ is the $k$-th Pad\'{e} pole in the upper half plane and $\frac{\eta_k}{\beta}$ is the corresponding coefficient~\cite{PhysRevB.87.085110}

\subsection{Decoherence and computational search for pointer states}
\label{subsec:dephasing}
The decoherence effect in this work is introduced in two different ways. One is based on a phenomenological model, which is called B\"{u}ttiker's probe (BP) model~\cite{PhysRevB.33.3020}.  The decoherence effect due to the B\"{u}ttiker's virtual probe is introduced by the virtual leads, represented by an additional line-width function $\Gamma_p$, where $p=1,2,\cdots, N$ with N being the total number of B\"{u}ttiker probes. The virtual probe is assigned to certain chemical potential, $\mu_p$.  As a consequence, another set of auxiliary density matrices $\varphi_p(t)$ has to be solved, which can be evaluated in the same way as $\varphi_\alpha(t)$. Since the B\"{u}ttiker's probes are introduced to mimic the decoherence effect, there should be no net current induced by the probes. Consequently, the chemical potentials of the probes, $\mu_p$, have to be adjusted to ensure that the net current through virtual probes is zero. In this way, all the electrons incident into the probes are scattered back to the system. As each virtual probe is a thermal equilibrium bath as a grand canonical of mixed electrons with random phase. Hence, the electrons scattered back to the system by the virtual probe have no phase relation to the incident electrons, i.e., the phase coherence is destroyed when electrons are scattered back to the system by the virtual probe. The phase incoherence between the incoming and outcoming electron from the virtual probe introduces the decoherence effects. Since the incoming and outcoming electron flux through the virtual probe is determined $\Gamma_p$, the strength of decoherence can be controlled by tuning $\Gamma_p$. 

Another way to introduce decoherence is to explicitly include the scattering mechanism. In particular, electron-phonon interaction is considered. The electron-phonon interaction is one of the major sources of decoherence in realistic systems. In the presence of phonon, the electron has the probability of being scattered off inelastically by phonons. The inelastic scattering induced by phonon introduces the phase-breaking mechanism. In this work, the electron-phonon interaction is taken into account in the polaron picture, which gets rid of the direct electron-phonon coupling. The corresponding EOM of density matrix was developed ~\cite{jcp4918771}. The electron-phonon coupling strength characterizes the probability of electrons being scattered inelastically by phonon. Hence, in this method, the decoherence effect can be tuned by the variation of electron-phonon coupling strength.

\begin{figure}
  \includegraphics[width=0.4\textwidth]{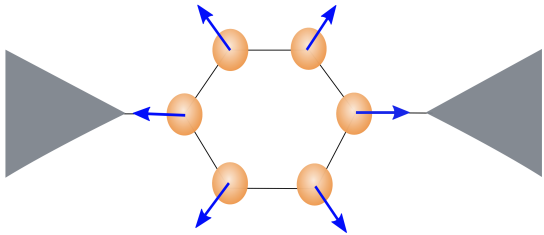}
  \vspace{-10pt}
  \caption{Illustration of benzene molecule connected to two electrodes and breathing phonon mode (blue arrow) of a benzene molecule.}
  \label{figbenzene}
\end{figure}

In the next section, the two models are used to examine the preferred states of an open electronic system which is a benzene molecule coupled to two leads as illustrated by Fig.~\ref{figbenzene} If preferred states exist, the density matrix will gradually become diagonal in the presentation the preferred states in the longtime limit. From another point of view, the density matrix can always be diagonal in its own eigenvectors. Hence, computational tracking the eigenstates of the density matrix can clearly show whether or not a well-defined set of pointer states can emerge from the decoherence dynamics. If the eigenstates of the density matrix are approaching a fixed set of states after a certain time scale (determined by the decoherence rate), then this fixed set of states can be defined as pointer states. The pointer states with different decoherence strength will be studied in the next section. 

\section{Results and Discussions}
\label{sec:result}

\subsection{Decoherence and pointer states selected by B\"{u}ttiker's virtual probe}
\label{sec:bp}
In this model, the decoherence effect is taken into account within the B\"{u}ttiker's virtual probe model as described in Sec.~\ref{subsec:dephasing}. In this study, each site of the benzene molecule is connected to an additional virtual probe. Hence, there are 6 virtual probes in total. For simplification, each site of the benzene molecule has the same coupling strength to the corresponding Buttiker's probe. 

In the following simulations, the coupling between two leads and the molecule is set as $\Gamma_{L}=\Gamma_R=2.0$~eV. Temperature is $300$~K and the bias voltage is $10$~meV. The bias voltage is switched on instantaneously in the following simulations. and the effect of bias voltage on the electronic structure of the molecule is described by a linear drop from the source to drain. The benzene molecule is described by the nearest-neighbor tight-binding model. The onsite energy is $0$~eV and the hopping integral between two neighbor sites is set as $2.0$~eV~\cite{solomon2010}. 

\begin{figure}[htb!]
  \includegraphics[width=0.45\textwidth]{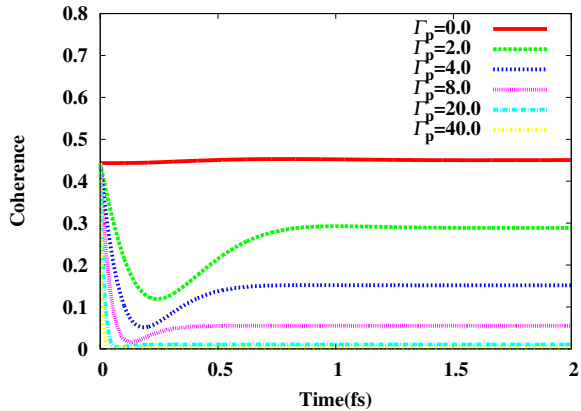}
  \vspace{-10pt}
  \caption{Dynamics of decoherence against different strength of virtual probe. The eigenstate of the density matrix in the strong strength limit is chosen as a representation to identify the coherence. Increasing the strength of the virtual probe reduces the coherence as a result of strong decoherence effect induced by the virtual probe.}
  \label{fig:virtualprobe_time}
  \vspace{-10pt}
\end{figure}

Initially, the system is in its thermal equilibrium state. After turning on the bias voltage, electrons
transport coherently from the source to the drain. The inclusion of B\"{u}ttiker's probe introduces
the decoherence effect. To examine the dynamics of decoherence, a certain basis set is chosen, denoted by $\{\phi_i\}$, which is the eigenstates of the density matrix in the strong coupling limit,  i.e., $\Gamma_p\rightarrow \infty$. The reason of choosing this representation will be explained later. Thus, the coherence of the system can be numerically tracked in the $\{\phi_i\}$ representation, which is defined as the summation of all the off-diagonal elements in the representation. The evolution of coherence is shown in Fig.~\ref{fig:virtualprobe_time}. Compared to $\Gamma_p=0$, it clearly demonstrates that the inclusion of B\"{u}ttiker's probe reduces the coherence. As described in Sec.~\ref{subsec:dephasing}, $\Gamma_p$ controls the decoherence strength, which is confirmed by the numerical simulations as illustrated in Fig.~\ref{fig:virtualprobe_time}. It is found that that the decoherence effect is enhanced by increasing $\Gamma_p$ and the coherence is reduced to zero in the strong coupling limit. Besides, the decoherence rate is enhanced with increasing $\Gamma_p$. If we track the eigenstates of the density matrix in real time under different $\Gamma_p$, all those eigenstates approach a fixed set of states after certain timescale (determined by the decoherence rate). That is, for the system studied here, pointer states can emerge in arbitrary $\Gamma_p$. Moreover, by tracking the eigenstates of the density matrix, it is found that the eigenstates of the density matrix also approach a fixed set of states with increasing $\Gamma_p$. These states are found to be the eigenstates of the system. But there are some differences: the system Hamiltonian has degenerate states, while the density matrix at strong coupling limit has nondegenerate states. Hence, even though the pointer states deform to the eigenstates of the system, only a specific set of eigenstates are selected out while all the superposed states are completely destroyed in the strong coupling limit, resulting in vanish of quantumness. This is also the reason why the eigenstates of the density matrix in the strong coupling limit are chosen as basis sets to characterize the coherence.

\subsection{Decoherence and pointer states selected by electron-phonon interaction}
As the weak coupling isn't able to destroy the coherence between states fully, the electron-phonon interaction is taken into account in the strong coupling regime. In this section, we show how does the electron-phonon interaction destroy the coherence and select the pointer states. 

Because this work focuses on the decoherence effects induced by electron-phonon interaction, only a single vibrational mode of benzene molecule is considered for simplification which is enough to gain insight into the problem. In particular, the breathing mode of the benzene molecule as illustrated by the blue arrow in Fig.~\ref{figbenzene} is focused. The interaction between electron and breathing phonon mode is described by the Hamiltonian
\begin{equation}
H_{ep}=g\sum_\mu (b^\dag+b)(c^\dag_\mu c_{\mu+1}+\text{H.c.}),
\end{equation}
where $g$ is the electron-phonon coupling strength and $b$ ($b^\dag$) denotes the annihilation (creation) operator of phonon. The Hamiltonian of phonon is described by $H_{p}=\omega b^\dag b$ with $\omega$ being the phonon frequency. Here $g$ is considered as a variable, and we show how does increasing $g$ select the pointer states. Since the onsite  energy of the benzene molecule is set to be $0$~eV, the Hamiltonian of the molecule is described by 
\begin{equation}
H_M=\sum_\mu t (c^\dag_\mu c_{\mu+1}+\text{h.c.}).
\end{equation}
It can be easily verified that $H_M$ commutes with $H_{ep}$. Consequently, $H_{ep}$ can be diagonalized in the representation of the eigenstates of $H_M$. Hence, in this section, the numerical simulations are carried on the representation of the eigenstates of the molecule. To identify the decoherence effect induced by the electron-phonon interaction, the density matrix is diagonalized and, an angle, $\theta$, between eigenstates of the density matrix and $H_M$ is used to characterize the coherence. $\theta$ denotes the angle of eigenstates of density matrix on the Bloch space to be rotated to reach the eigenstates of $H_M$. In the following simulations, the frequency of the breathing mode is set as $100$~meV, and the phonon is assumed in thermal equilibrium state associated with temperature $300$~K.

\begin{figure}[htb!]
  \includegraphics[width=0.4\textwidth]{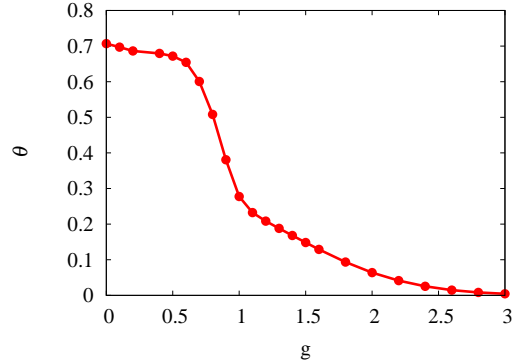}
  \vspace{-10pt}
  \caption{Angle to be rotated on the Bloch space to reach eigenstates of $H_M$ as function of electron-phonon coupling strength. The frequency of the breathing mode is set as 100~meV and phonon is assumed in an equilibrium state with room temperature.}
    \label{fig:angle_g}
\end{figure}

Fig.~\ref{fig:angle_g} plots the angle to be rotated on the Bloch space to reach the eigenstates of $H_M$ with respect to different electron-phonon coupling strength. As shown in the figure, the angle decreases with increasing electron-phonon coupling strength. Which means that the increasing electron-phonon interaction induces more significant phase-breaking between the eigenstates of $H_M$. In the strong coupling limit, the $\theta$ reduces to 0, which indicates that the eigenstates of $H_M$ are selected out with increasing electron-phonon coupling strength. In other words, the eigenstates of $H_M$ forms the pointer states of the system, which coincides the result of Sec.~\ref{sec:bp}.

In the strong coupling limit, pointer states deform to the eigenstates of the system, which indicates that the quantumness is totally suppressed. In general, the statistic of an open electronic system is noncanonical. While the noncanonical statistics vanishes when the eigenstates of density matrix coincide with the system's energy eigenstates, which indicates the quantumness of the open system vanishes as well~\cite{PhysRevE.86.021109}. The deformation to the eigenstates of $H_M$ in the strong coupling limit can be easily understood in the present model. As when $g\rightarrow \infty$, the Hamiltonian of the molecule-lead coupling can be neglected and the total Hamiltonian becomes
\begin{equation}
H_T\approx H_M + H_p + H_{ep}
\end{equation}
with $[H_M,H_{ep}]=0$. In this case, it is well known that the coherence between the eigenstates can be completely destroyed after the decoherence process~\cite{breuer2006theory}. Consequently, the eigenstates of the system become the pointer states in this limit.

It should be noted that the strengths of the virtual probe or the electron-phonon coupling are too strong to be practical. Since our molecule is small, the quantum effect is significant. The realistic strengths cannot destroy the coherence completely. These ``unphysical" parameters are adopted to demonstrate the selection of pointer states. In realistic electronic devices, the size of interest is usually much larger than that of benzene. The longer lengthscale of the electrons transporting through the device increases the possibility of being scattered, thus effectively introduces stronger decoherence effect and ultimately destroy the coherence and reach the classical limits. As a result, the electron transport mechanism transit from the ballistic transport to Ohm's law.

\section{Summary}\label{sec:summary}
In summary, pointer states of an open electronic system under non-equilibrium condition is investigated via two different models. The pointer states are computationally defined by tracking the variation of eigenstates of the density matrix in real time. It is found that pointer states can emerge in different coupling regime. The decoherence induced by the virtual probe or electron-phonon coupling destroys the superposition of states after a certain time scale which is determined by the decoherence rate. As a consequence, a coupling-strength dependent set of pointer states is selected after the decoherence process. Moreover, the pointer states deform to the eigenstates of the system with increasing coupling strength for both the two models.

\begin{acknowledgments}
H.J was supported by the National Natural Science Foundation of China (NSFC, Grants No.11334015). The work at Los Alamos National Laboratory (LANL) was supported by the LANL Directed Research and Development Funds (LDRD). LANL is operated by Triad National Security, LLC, for the National Nuclear Security Administration of the U.S. Department of Energy (Contract No. 89233218NCA000001). 
\end{acknowledgments}

\bibliography{ref}
\end{document}